# Accurate theoretical fits to laser ARPES EDCs in the normal phase of cuprate superconductors

Philip A. Casey[1], J. D. Koralek[2,3], D.S. Dessau[2,3] and Philip W. Anderson[1*]

[1] Department of Physics, Princeton University, Princeton, New Jersey 08544, USA
[2] Department of Physics, University of Colorado, Boulder, Colorado 80309, USA
[3] JILA, University of Colorado and NIST, Boulder, Colorado 80309, USA
* pwa@princeton.edu

Anderson has recently proposed a theory of the strange metal state above $T_c$ in the high $T_c$ superconductors.[1] It is based on the idea that the unusual transport properties and spectral functions are caused by the strong Mott-Hubbard interactions and can be computed by using the formal apparatus of Gutzwiller projection. In ref. 1 Anderson computed only the tunneling spectrum and the power-law exponent of the infrared conductivity. He had calculated the energy distribution curves (EDCs) in angle resolved photoemission spectroscopy (ARPES) but was discouraged when these differed radically from the best ARPES measurements available at the time, and did not include them. In this letter we compare the spectral functions computed within Anderson's model to the novel laser-ARPES data of Dessau's group.[2,3] These are found to capture the shape of the experimental EDCs with unprecedented accuracy and in principle have only one free parameter.

Recently, ARPES measurements using 6 eV laser radiation have become available.[2,3] The relatively low photon energy in these experiments greatly increases momentum resolution relative to higher energy ARPES, reducing the energy width of the EDCs for dispersive electronic states. Sensitivity to the bulk-physics is also increased with laser ARPES, resulting from longer photoelectron mean-free paths in the solid at low kinetic energy. Historically, accurate fitting of ARPES EDCs has been complicated by poor energy and momentum resolution as well as the existence of a large extrinsic background.[4,5] However, these effects are minimized when low energy photons are used. These improvements have allowed for the most accurate measurements of the cuprate EDC lineshape to date.[2]

In ref. 1 it was shown that the effect of the Gutzwiller projection in the idealized case of a simple Fermi surface at absolute zero is to multiply the free-particle Green's function in space-time by a factor $t^{-p}$, where p is $\frac{1}{4}(1-x)^2$ and x is the hole doping level. This value of the exponent is approximately confirmed by the exponent of the infrared conductivity dependence on frequency.[6] Fig. 1 shows the agreement of this exponent with Anderson's theory. Motion of a particle near the Fermi surface is essentially one-dimensional, so we may take the free particle Green's function in space-time as $1/(x-v_F t)$. In order to get the imaginary part of G (the density) in k and frequency space, we must Fourier transform G(x,t),

$$G(k,\omega) = \iint dx\, dt\, e^{i(kx-\omega t)}\, t^{-p} / (x - v_F t) \qquad (1)$$

Doing the x integration by a contour integration (the sign of t determines which way to close the contour), this becomes,

$$G(k,\omega) = \int dt\, t^{-p}\, e^{i(v_F k - \omega)t} \propto (v_F k - \omega)^{-1+p} \qquad (2)$$

The imaginary part of this expression is the T = 0 EDC. If p = 0, this is just a delta function at the quasiparticle energy, $v_F k$, but if p is finite it has an imaginary part for all $\omega > v_F k$. The quasiparticle becomes a cut singularity, not a pole, in the complex plane and does not have a finite residue *at* the singularity, i.e. it has quasiparticle residue Z = 0.

The absence of a finite Z has a profound effect on the temperature behavior. If there are ordinary quasiparticles their energies are not affected by thermal fluctuations. Impurity scattering, for instance, only changes their spatial wave functions - and there is no tendency towards "ω, T scaling" which one naively expects from the analogy between the Boltzmann distribution $e^{-E/k_B T}$ and the Schrodinger time dependence $e^{iHt}$. Another way to say it is that in that case there are conservation laws, effected by Ward identities, which restrict the scattering, while in the current case there are only phase space restrictions, so that the effect of raising the energy is just the same as raising the temperature. A heuristic approximation to the effect of finite T is to insert a relaxation rate $\Gamma = AT$, where A is a constant of order unity, so that we replace $t^{-p}$ by $t^{-p} e^{-\Gamma t}$. In ref. 7 Yuval suggested that for finite T, $t \rightarrow \sinh(\pi T t) / \pi T$, so that we might expect $A \sim \pi p$, but that is only an estimate. The excitations into which any decay takes place are Fermionic, so that we must also multiply by the Fermi function of energy hω/2π.

The observed spectra clearly show a broadening which increases with increasing binding energy. A conventional Fermi liquid would have the scattering go as $\omega^2$, producing additional decay for the curves with $k$ farther from $k_F$. In a supplement to be published for ref. 1, Anderson shows that there is an underlying, "hidden" Fermi liquid of excitations, which can therefore be expected to be scattered at the conventional rate $\propto \omega^2$. The final expression for the EDC then becomes,

$$Intensity = \operatorname{Im}\{G\} = \operatorname{Im}\{\frac{f(\omega/T)}{[(v_F k - \omega) + i\Gamma]^{1-p}}\}$$
$$= f(\omega/T) \frac{\sin[(1-p)\cot^{-1}([\omega - v_F k]/\Gamma)]}{[(\omega - v_F k)^2 + (\Gamma)^2]^{(1-p)/2}} \quad (3)$$

Here we have expressed all frequencies and temperatures in the same energy units, and $f$ is the Fermi distribution,

$$f = 1/(1 + e^{\hbar\omega/k_B T}) \quad (4)$$

The lineshape, Eq. (3), is just the Doniach-Sunjic lineshape[8] multiplied by the Fermi function. Though it shares with the simple power law the fact that the singularity is not a pole and has Z=0, it was historically often mistaken for a simple Lorentzian.

In principle it is possible to calculate the temperature dependence of the Green's function, and its Fourier transform $G(k,\omega)$, in a less heuristic way, since Yuval has actually given a prescription for the Green's function in the x-ray problem at finite T. However, the formalism would involve enormously complicated contour integration and we think the above heuristic form is adequate for the current purposes. In explicit form Eq. (3) is quite complicated, and it should be: the central portion resembles a Lorentzian, while the tail behaves like $\omega^{-1+p}$.

In Fig. 2, we demonstrate fits of Eq. (3) (plus a small, constant background – see figure caption) to ARPES EDCs at three k points and three temperatures from ref. 3. Many more k points were fit as seen in Fig. 3. We also compare the fits of Eq. (3) with Lorentzian fits similar to those given in ref. 3. The sample was nearly optimally doped $Bi_2Sr_2CaCu_2O_{8+\delta}$ with a transition temperature around 90 Kelvin, and *k* was along the nodal direction, so at the temperatures between 100 and 200 Kelvin there is little or no effect of the superconducting gap. The parameters for each curve are also given in Fig. 2.

The Doniach-Sunjic fits are clearly superior to the pure Lorentzian fits at capturing the asymmetry of the measured EDCs. In principle these new fits have at most one truly free parameter - an approximation leaves one numerical coefficient free within narrow limits. The current fits are not perfect, nor should they be - the exponential cutoff assumed is probably too slow, and we might expect a steeper falloff on the low-energy side. It is also noted that a much smaller background, B, is needed to fit with the Doniach-Sunjic lineshape than with the pure Lorentzians (compare $B_L$ to $B_A$), and this background is more in line with the expectations of simple inelastic scattering. Fermi liquid and Marginal Fermi liquid fits also can capture much of this asymmetry and will be discussed in detail in a future publication by J.D.K., *et al.*

Figure 3 shows the trends of the extracted relaxation rate, $\Gamma$, as a function of temperature and binding energy. A similar plot for the Lorentzian fits can be seen in Fig. 3b of ref. 3. The current fits show slighter smaller $\Gamma$ and a smoother behavior, which we parameterize as $\Gamma = A\, k_B T + C\, v_F^2 (k - k_F)^2$. As seen in Fig. 3, the extracted A and C values are nearly constant over all the individual EDC fits, implying a near-universal set of fits.

The spectral lineshapes predicted by the strange-metal theory of cuprate superconductors are found to fit the laser-ARPES EDCs accurately and with a minimal number of free parameters. Compared to simple Lorentzian fits, the asymmetry in the EDCs is well characterized and the backgrounds minimized and well controlled. It now should be possible, with considerably more computational effort, to analyze the curves in the superconducting state and understand the subtle intensity transfers which occur there.

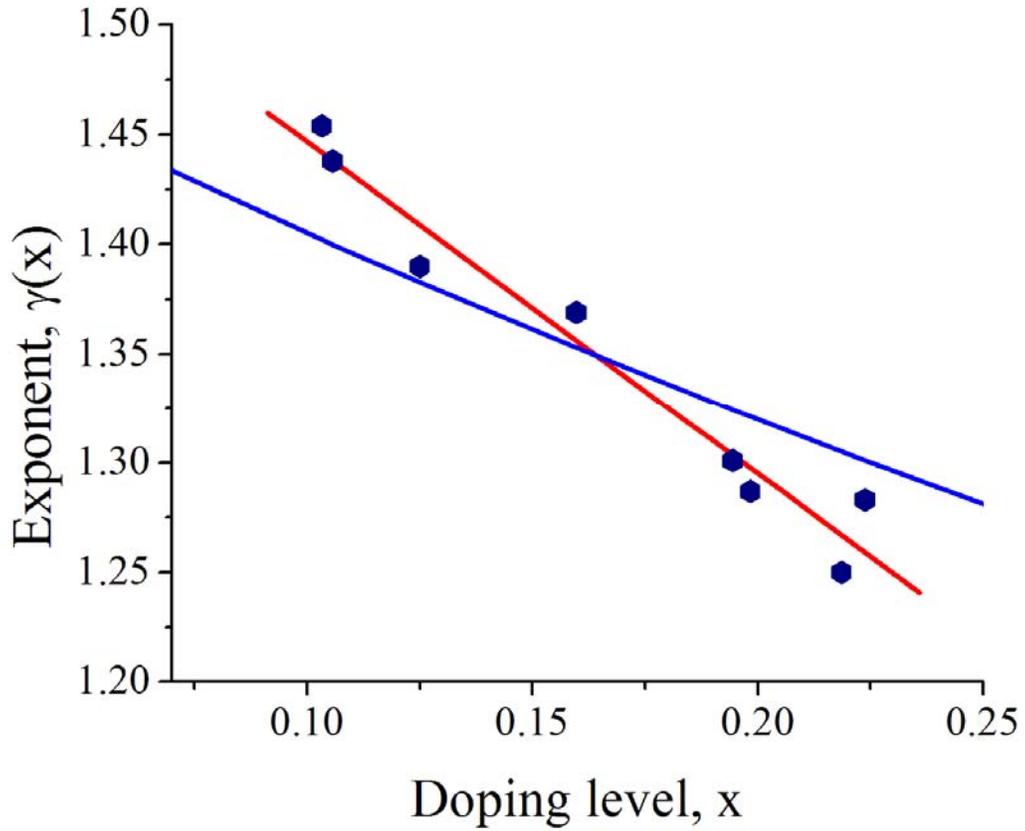

Figure 1. Infrared spectrum exponents for $Bi_2Sr_2CaCu_2O_{8+\delta}$. Data points from ref. 6 with linear best fit of ref. 6 (red line) and predicted value from ref. 1 (blue line). The predicted exponent stems from $\sigma(\omega) = (i\omega)^{-2+\gamma}$ with $\gamma = 1 + 2p$, and p is given in the text.

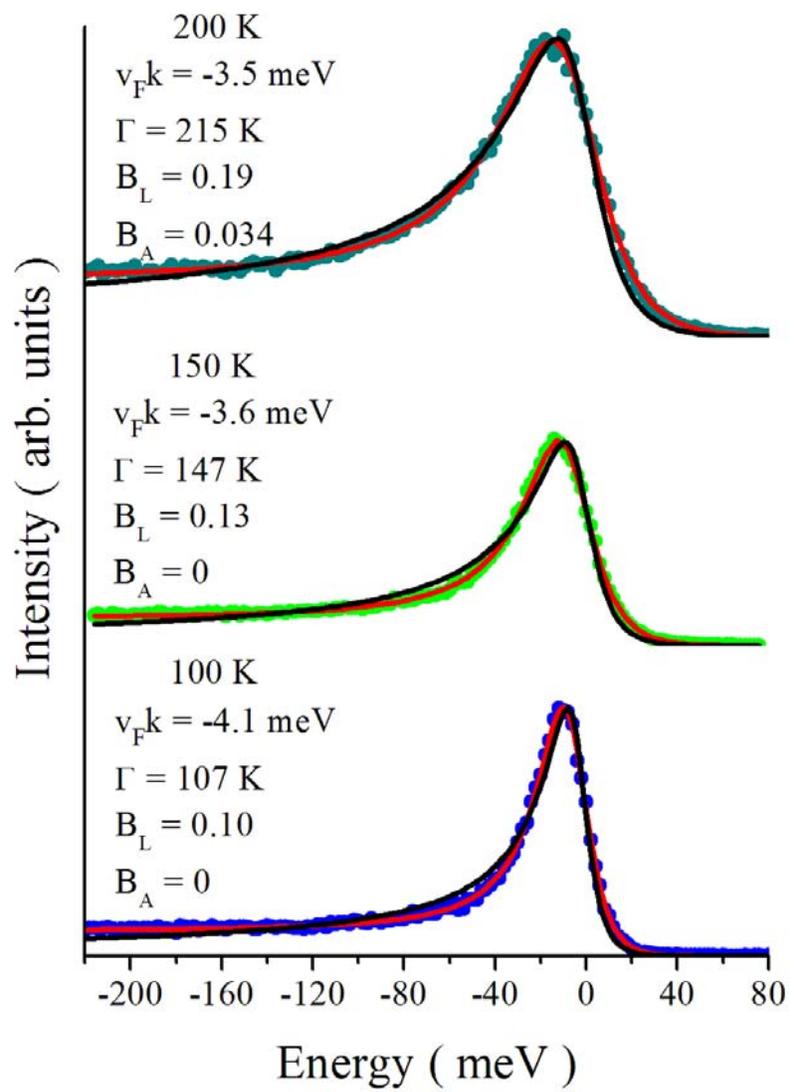

Figure 2a.

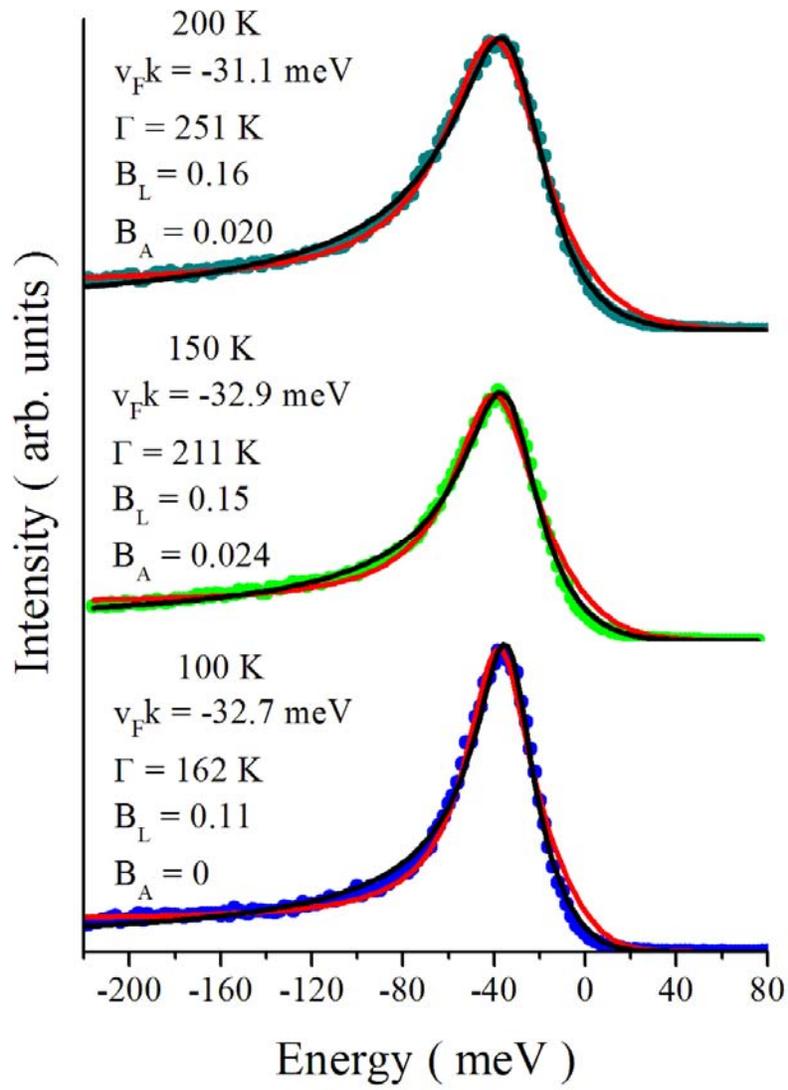

Figure 2b.

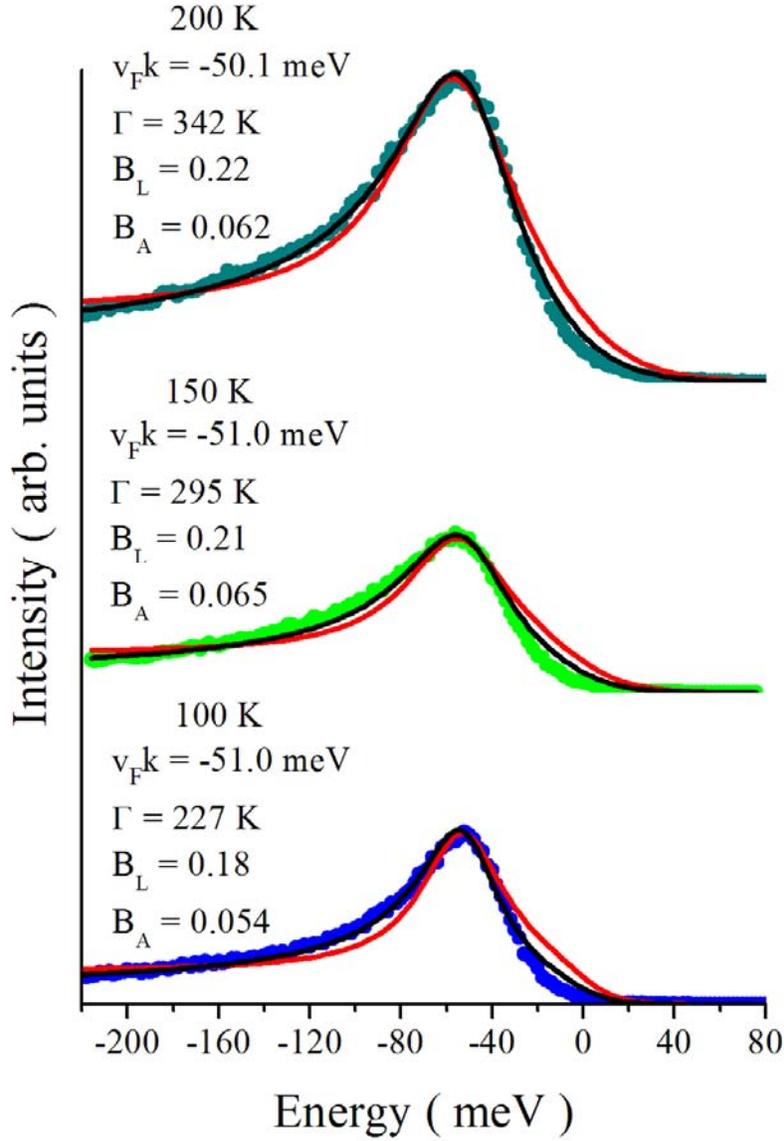

Figure 2c.

Figure 2. ARPES EDCs in the "strange metal" phase above $T_c$ of optimally doped $Bi_2Sr_2CaCu_2O_{8+\delta}$. a, k near $k_F$ and b,c at higher quasiparticle energies, as quoted. Points are experimental and red curves are fitted Lorentzians with background, $B_L f(\omega/T)$. Black curves are theoretical fits from present paper, Eq. (3) with background $B_A f(\omega/T)$. $B_A$ is found to be negligible: compare $B_L$ to $B_A$. Backgrounds are measured in units of the intensity relative to the peak of the EDC.

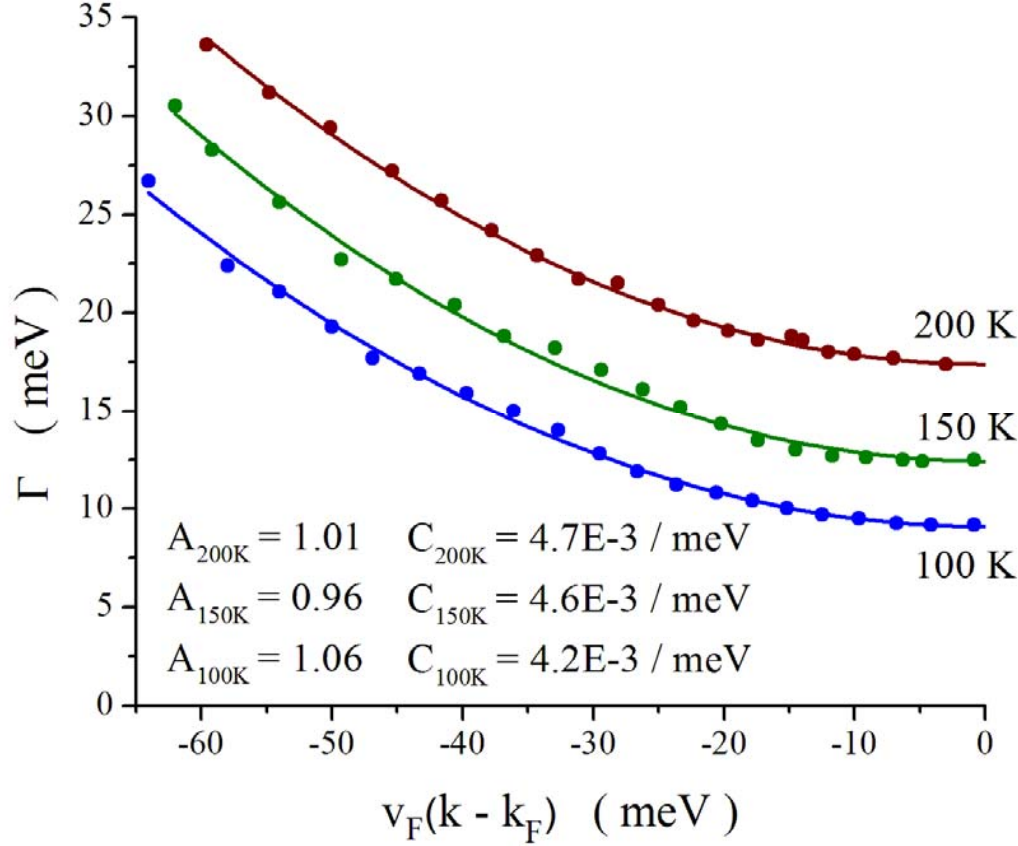

Figure 3. Relaxation rate, $\Gamma$, (data points) extracted by fitting the experimental EDCs. Fits were Eq. (3) plus a small, non-negative background, $B_A f(\omega/T)$. Solid lines are best fits of $\Gamma = Ak_BT + C v_F^2(k - k_F)^2$. The given empirical values fit well to universal parameters $A = 1$ and $C = 4.5 \times 10^{-3}$ / meV.

Acknowledgements

P.A.C. gratefully acknowledges support from a National Sciences and Engineering Research Council of Canada Post-Graduate Scholarship. Support for J.D.K and D.S.D. came from DOE Grant No. DE-FG02-03ER46066, with other funding from NSF Grant Nos. DMR 0706657, and by the NSF ERC for Extreme Ultraviolet Science and Technology under NSF Award No. 0310717. J.D.K. and D.S.D. thank J. F. Douglas, N. Plumb, Z. Sun, A. Fedorov, J. Griffith, S. Cundiff, M. Murnane, and H. Kapteyn for help with the experiments and Y. Aiura, H. Eisaki, and K. Oka for growing the Bi2212 samples.